\RequirePackage{fix-cm}
\documentclass[smallextended]{svjour3}       % onecolumn (second format)
\usepackage{amssymb, lipsum, color}
\smartqed  % flush right qed marks, e.g. at end of proof
\usepackage{graphicx}
\RequirePackage{amsmath} 
\RequirePackage{amssymb} 
\RequirePackage{latexsym} 
\RequirePackage[mathscr]{eucal} 
\RequirePackage{verbatim} 
\RequirePackage{enumerate} 
\RequirePackage{xspace}
\RequirePackage[colorlinks]{hyperref} 
\RequirePackage[all]{xy}
\RequirePackage{color}
\RequirePackage{geometry}                % See geometry.pdf for layout options. 
\RequirePackage{graphicx}
\RequirePackage{xspace}

\newcommand{\mc}[1]{\mathcal{ #1}}

\newcommand{\mbb}[1]{\mathbb{ #1}}

\newcommand{\nel}{\newline\noindent}
\newcommand{\ket}[1]{|{ #1}\rangle}
\newcommand{\bra}[1]{\langle{ #1}|}
\newcommand{\prob}[3]{\mathtt{Pr}_{ #1}^{({#2}^{#3})}}
\newcommand{\proj}[3]{{P}_{ #1}^{(#2^{#3})}}
\newcommand{\tensr}[2]{{\otimes}^{ #1}{\mathbb C^{#2}}}
\newcommand{\tr}{\mathtt {Tr}}
\begin{document}

\title{A note on many valued quantum computational logics\thanks{The authors acknowledges the support of MIUR within the FIRB project ``Structures and Dynamics of Knowledge
24 and Cognition", Cagliari: F21J12000140001 and of RAS within the project ``Modeling the uncertainty: Quantum Theory and Imaging Processing".}
}


\author{Giuseppe Sergioli         \and
        Antonio Ledda}


\institute{A. Ledda, G. Sergioli \at
              Universit\`a di Cagliari, Italia \\
              Tel.: +39-070-6757120\\
              \email{antonio.ledda@unica.it}				\\
              \email{giuseppe.sergioli@gmail.com}             
}

\date{Received: date / Accepted: date}

\maketitle

\begin{abstract}
The standard theory of quantum computation relies on the idea that the basic information quantity is represented by a superposition of elements of the canonical basis and the notion of probability naturally follows from the Born rule. 

\noindent
In this work we consider  three valued quantum computational logics. More specifically, we will focus on the Hilbert space $\mathbb C^{3}$, we discuss extensions of several gates to this space and, using the notion of effect probability, we provide a characterization of its states.
\end{abstract}

\section{Introduction}
\label{intro}
The usual notion of \emph{uncertainty} seems to be tightly related to an epistemic condition \cite{Lind}. A typical case: a coin was flipped, but a specified knower couldn't see which side of the coin faced up when it landed. It seems to be generally accepted that uncertainty deals with ignorance: a certain predicate is uncertain, with respect to a specific knower, if the information available is not sufficient to determine its applicability.

On the other hand, \emph{vagueness} seems to be unrelated with ignorance. Instead, this notion often refers to concepts whose extensions are lacking in clarity \cite{Keefe2000}, \cite{Fi}, \cite{Williamson 94}. Natural examples of vague concepts are predicates admitting  ``border-line cases'', in which it is hard to sharply determine whether an object falls completely in the extension of the predicate or not. Rather, an object may possess specific properties to some extent. A successful framework for dealing with vagueness is provided by \emph{many-valued logics} \cite{Goguen 69}, \cite{Ha}.

In  the microscopic domain, appreciable overlaps between the concepts of uncertainty and vagueness are to mention.
A remarkable example is the ``Stern-Gerlach experiment'' that shows how particles possess an intrinsic angular momentum that can assume certain discrete values only. The experiment is normally conducted using atoms or  electrically neutral particles, that are treated as classical spinning dipoles; they will precess in a magnetic field because of the torque that the field exerts on the dipoles. If the magnetic field where the particles move through is not homogeneous, then the force on one end of the dipole of each particle will be slightly greater than the opposing force on the other end, so that there is a net force which deflects the trajectory of the particles. If the particles were classical spinning objects, one would expect that the distribution of their spin angular momentum vectors will be random and the spectrum will be continuous, because each particle will be deflected by a different amount, producing a density distribution on the detector screen. Instead, the particles passing through the Stern$-$Gerlach apparatus are deflected either up (say $\ket 0$) or down (say $\ket 1$) by a specific amount, because - how is well know nowadays -  the spectrum of quantum angular momentum is discrete.

Formally speaking and by using the common Dirac notation, we describe the state of a particle that is passing through a Stern$-$Gerlach apparatus as the following superposition

$\ket\psi=a\ket0+b\ket 1$,
where $|a|^2$ and $|b|^2$ are the probabilities to detect the particle with spin up or down, respectively. 
The state $\ket\psi$ is a \emph{pure state}, that represents a maximal piece of information that cannot be increased by any further observation. However, once a state $\ket\psi$ is fixed, by its very nature, it involves an amount of uncertainty, which in this case is a property of the state not related to the observer's knowledge. 
A state may also be \emph{mixed} and it can represent a non-maximal piece of information, mathematically represented by a density operator. 

We will see that, at the microscopic level, uncertainty and vagueness can be captured under several degrees of freedom rendering the two notions amenable of interactions not available in the classical world.
%several possible combinations (and meanings) of vagueness and uncertainty can be singled out.
%%%%%%%%%%%%%%%%%%%%

Useful tools for inquiring into possible interplays of vagueness and uncertainty in the quantum realm are provided by \emph{quantum computational logics}. These logics, investigated by Maria Luisa Dalla Chiara,
Roberto Giuntini and other authors, including, for example, the present papers \cite{BL}, \cite{CDGL}, \cite{CDGL04}, \cite{DGG}, \cite{DGL05}, \cite{GLSP}, differ from the well known Birkhoff-von Neumann approach to quantum logic, where propositions ascribing properties are represented by projection operators -- or, equivalently, by closed subspaces of a Hilbert space. In quantum computational logics meanings of sentences are no longer represented by projector operators on a Hilbert space, but by means of \emph{quantum
information quantities}: qubits, qutrits, density operators. In this framework fuzzy-like structures appear at different levels, and with different status. The aim of this paper is to discuss this further bridge between many-valued and quantum logics. 

The paper is organized as follows: in Section \ref{sec:2} we provide all the basic notions necessary to render the article self-contained; in Section \ref{sec:3} we focus on the concepts of \emph{uncertainty}, \emph{mixedness} and \emph{degrees of truth} as a foundational motivation in the context of many-valued quantum computational logics; in Section \ref{sec:4} we discuss some unary quantum gates in the context of standard and many-valued quantum computational logics; in  Section \ref{sec:5} we introduce the notion of \emph{effect probability}. This notion will be expedient in showing a characterization of mixed states in $\mathbb C^3$. Finally, we close the paper with some comments.
\section{Basic Notions}\label{sec:2}
For the reader's convenience, we recall in this section all the basic notions required for a complete understanding of the paper.

Consider the $n$-fold tensor product Hilbert space $\otimes^n\mathbb C^d$, with $n\ge1$ and $d\ge2$.

The canonical orthonormal basis $\mathcal B^{(d^n)}$ of $\otimes^n\mathbb C^d$ is defined as follows:
$$\mathcal B^{(d^n)}=\{\ket{x_1,...,x_n}:x_i\in\{0,\frac{1}{d-1},\frac{2}{d-1},...,1\},\forall i\in\{1,...,n\}\}$$ where 
\begin{itemize}
\item $\ket {x_1,...,x_n}$ is an abbreviation for the tensor product $\ket{x_1}\otimes...\otimes\ket{x_n};$
\item the vector $\ket{\frac{i}{d-1}}\in\mathbb C^d$ (with $0\le i\le d-1$) is a $d$-dimensional column vector with $1$ in the $(i+1)^{th}$-entry and $0$ in all the other $d-1$ entries.
\end{itemize} 

\begin{definition}Qudit

A unit vector in the Hilbert space $\mathbb C^d$ (with $d\ge 2$) is a \emph{qudit}. As a special case, if $d=2$ the unit vector is the so called \emph{qubit}, whose extensive expression $\ket\psi=a\ket 0 +b \ket 1$  (where $|a|^2+|b|^2=1$) was already mentioned in the previous Section.
\end{definition}
\begin{definition}{Quregister and Qumix}

A \emph{quregister} is a unit vector in $\otimes^n\mathbb C^d$ and a \emph{qumix} (or \emph{mixed state}) is a density operator in $\otimes^n\mathbb C^d$.
\end{definition}

So, a vector in $\otimes^n\mathbb C^d$ is a $n$-fold tensor product of $d$-dimensional vectors. Trivially, qudits are special cases of quregisters.

\begin{definition}{Truth-values of a quregister}\label{tv}

We say that the truth-value of a quregister $\ket{x_1,...,x_{n-1},\frac{i}{d-1}}\in \otimes^n\mathbb C^d$ 
is $\frac{i}{d-1}$, with

$0 \le i  \le d-1$. 
\end{definition}

For, the truth-value of a quregister only depends on its last component. In particular, if $i=0$ we say that the register is \emph{false} and if $i=d-1$ we say that the register is \emph{true}.

Let us remark that the number of different truth-values over the Hilbert space $\otimes^n\mathbb C^d$ is $d$, for any value of $n$. 

\begin{definition}{The truth-value projectors}

A \emph{truth-value projection} on $\otimes^n\mathbb C^d$ is a projector $P^{(d^{n})}_{\frac{i}{d-1}}$ whose range is the closed subspace spanned by the set of all quregisters whose $n^{th}$ component is $\ket {\frac{i}{d-1}}$, where $P^{(d^{n})}_{\frac{i}{d-1}}=I^{(n-1)}\otimes P^{(d)}_{\frac{i}{d-1}}$ and $0 \le i \le d-1.$% \commento{Proporrei la notazione: $P^{(d^{n})}_{\frac{i}{d-1}}$, in tal modo risolviamo anche l'ambiguita!}

In particular, the \emph{truth-projection} of $\otimes^n\mathbb C^d$ is the projection operator $P^{(d^{n})}_1$ whose range is the closed subspace spanned by the set of all true quregisters of $\otimes^n\mathbb C^d$.
\end{definition}

As an example, let us note that the projector operators $P^{(2^{2})}_1\in \otimes^2\mathbb C^2$ and $P^{(4)}_1\in\mathbb C^4$, take, respectively, the form:

$P^{(2^2)}_1=I\otimes P^{(2)}_1=\left(\begin{matrix}
0&0&0&0\\
0&1&0&0\\
0&0&0&0\\
0&0&0&1
\end{matrix}
\right)$ and  $P^{(4)}_1=\left(\begin{matrix}
0 &0    & 0 & 0\\
0&0&0&0\\
0&0&0&0\\
0&0&0&1
\end{matrix}
\right).$
%\footnote{Let us note that, in general, a projector $P^{(n)}_{\frac{i}{d-1}}$ may refer to different Hilbert spaces. However, the space we are refering to will be always evident from the context.} 

For any qumix $\rho\in\otimes^n\mathbb C^d$ it is possible to introduce a notion of probability, according with the Born rule, as follows:

\begin{definition}$\frac{i}{d-1}-probability$

Let $\rho$ be a qumix in $\otimes^n\mathbb C^d$. The probability \cite{DGS} that $\rho$ has the truth-value $\frac{i}{d-1}$ (with 

$0 \le i \le d-1$) is defined by as: $$\mathtt{Pr}^{(d)}_{\frac{i}{d-1}}=\mathtt{tr}(P^{(d^n)}_\frac{i}{d-1}\rho),$$
where $\mathtt {tr}$ is the trace functional.
\end{definition}
From an intuitive point of view, $\mathtt{Pr}^{(d)}_{\frac{i}{d-1}}$ represents the probability that the information stored by the qumix $\rho$ is the truth-value $\frac{i}{d-1}.$

The unitary evolution of quregisters and qumixes is dictated by quantum logical gates. \emph{Quantum logical gates} (and the quantum operations they naturally induce) are unitary transformations that map quregisters (qumixes) in $\otimes^n\mathbb C^d$ into quregisters (qumixes) in $\otimes^n\mathbb C^d$.
 From a foundational perspective, it could be useful to distinguish between
\begin{itemize}
\item{semiclassical quantum gates: unitary operators that transform basis elements into basis elements;}
\item{genuinely quantum gates: unitary operators that transform basis elements into superposition states.}
\end{itemize}

In the rest of the paper, we will mostly be interested in the Hilbert space $\mathbb C^3$. We say that a unit vector in $\mathbb C^3$ is a \emph{qutrit}, a density operator (qumix) in $\mathbb C^3$ is a \emph{qutrit-density operator} and a quantum logical gate on $\mathbb C^3$ is a \emph{qutrit-gate}.
%\begin{itemize}
%\item pure states
%\item density operators
%\item probabilities in qc logics
%\item notion of truth: the target information
%\item qubits - basis
%
%
%\end{itemize}

\section{Uncertainty and degrees of truth}\label{sec:3}

\subsection{Uncertainty}
The concept of uncertainty has been a major topic of discussion for engineers, philosophers and mathematicians working on statistical theories. In particular, in the context of risk analysis \cite{CM}, many scholars categorize uncertainties into two main families: aleatoric and epistemic uncertainty \cite{DD}, \cite{Matt}. The first family groups those deriving from randomness in samples while the second those that stem from a lack of knowledge.
Examples of aleatoric uncertainty are the occurrence of an hurricane in the Gulf of Mexico, or the hight of an arbitrary individual in a certain population. On the other hand examples of epistemic uncertainty include, for instance, the global effect clouds formation have on the temperature of earth, or the nature of certain earthquake mechanisms \cite{PC}.
Aleatoric uncertainty comprises unknowns that vary each time the same experiment is performed.
According with \cite{DD}, aleatoric uncertainty is related to the intrinsic randomness of a phenomenon, and  epistemic uncertainty is caused by a lack of knowledge.

In both cases, however, the notion of uncertainty is tightly related to a condition of the modeler. It is a concept that has to deal with \emph{ignorance}: a given predicate is uncertain, relative to a specific modeler, when the  available information is not sufficient to determine its applicability \cite{Lind}.
However, once we enter the quantum domain, the concept of \emph{uncertainty} assumes a different status, and new degrees of freedom will be available for this concept. Let us recall for a moment the classical two-slits experiment. A coherent light source is placed in front of a screen that contains two parallel slits. The wave nature of light causes the light waves passing through the two slits interfere, producing bright and dark bands on the screen placed behind the slits. The interference phenomenon is formally expressed by a superposition state $\ket\psi=a\ket0+b\ket 1$
where $\ket \psi$ represents the state of a photon before coming up against one of the two slits, $\ket 0$ and $\ket 1$ represent the states of a photon after it passed through either the first or the second slits, respectively, and $|a|^2$ and $|b|^2$ are the respective probabilities. 
Differently from the classical case, the state $\ket\psi$ is ontologically superposed: before a measurement occurs, the photon is neither in the state $\ket0$ nor in $\ket1$ but in a mere superposition between both. This feature of $\ket \psi$ is unrelated to the ignorance of the modeler and this fact is formally expressed by the unitarity of the state $\ket \psi$, in fact $\ket\psi$ is said to be a \emph{pure} state.
The concepts of \emph{complete knowledge} (or maximal information), \emph{unit vector} and \emph{pure state} are, in this context, one and the same notion.

From a quantum logical point of view also, once we agree on a logical basis, uncertainty is related to proper features of the state only, without any reference to any knowledge of a possible observer  \cite{DGL}.
Upon setting $\ket0$ and $\ket1$ to be our choices for the truth values ``false" and ``true" respectively, the superposed state $\ket\psi=a\ket0+b\ket 1$ expresses a \emph{logical uncertainty} (from now on simply \emph{uncertainty}) between the possible truth values: $\ket 0$ and $\ket 1$, with the respective truth-probabilities $|a|^2$ and $|b|^2$ (with $|a|^2+|b|^2=1$), in accord with the Born rule.

However, it is not the case that, in general, any superposition state has non-zero truth-probability. Indeed, consider the state $\ket{\psi'}\in\otimes^2\mathbb C^2$:

\begin{equation}\label{eq:superposed}
\ket{\psi'}=\frac{1}{\sqrt 2}\ket{00}+\frac{1}{\sqrt 2}\ket{10}=(\frac{1}{\sqrt 2}\ket{0}+\frac{1}{\sqrt 2}\ket{1})\otimes\ket0
.
\end{equation}
It can be easily verified that $\ket{\psi'}$ is a superposition state whose truth-probability is $0$.
On the other hand (due to the non-commutativity of the tensor product), the state $\ket{\psi''}=\ket0\otimes(\frac{1}{\sqrt 2}\ket{0}+\frac{1}{\sqrt 2}\ket{1})$ is both an uncertain and superposed state in $\otimes^2\mathbb  C^2$.

%\noindent Perhaps, the main outcome of these observations can be rendered as follows: once we stick on a logical base, the uncertainty is a property of the state. As an example, consider the case of $\mbb C^2$. If we agree on the notions of ``true and false states'' (in case of $\mbb C^2$, $\ket0$ and $\ket1$, respectively), for a state $\ket \psi$ the notion of uncertainty is \emph{ontological}. \\
% Actually, it depends only on internal properties of $\ket\psi$: contrarily to the classical case, it does not rely on any epistemic condition.
 
%\begin{equation}

%\end{equation}

\subsection{Mixedness}\label{M}
The information a given system provides can be considered along an alternative degree of freedom: its \emph{mixedness}.

In general, a quantum system may not be in a pure state, and the information quantity it describes may not be maximal. This could be due to, e.g., a non-complete efficiency of a preparation procedure, or, in general, to the interaction of the system with the environment, so that decoherence phenomena may arise corrupting the experimenter's knowledge on the system.
On the other hand, there are interesting processes that cannot be represented by unitary evolutions.
For instance, consider the case when, at the end of a computational process, a non-unitary operation -- a measurement -- is applied, and the state of the system collapses into a probability distribution over pure states, namely \emph{a proper mixed state} (mixed state for short) \cite{TB}.
A  mixed state represents a \emph{non-maximal information} on the system, that could be increased by further observations.  
In this case an evident epistemic feature comes into play. However, even if the concept of mixed state patently involves an observer, the property of ``being mixed'' should not be regarded as an exclusively epistemic condition devoid of any ontological commitment. In the microscopic context, indeed, any observation substantially modifies a state; and the property of  ``needing further observation to be completely known'' should be considered an ontological feature of a system. As an example, quantum decoherence phenomena \cite{Z} consist in the loss of the coherence of the phase angles between the components of a system in a quantum superposition: an amount of information from the system vanishes into the environment (in accord with a sort of \emph{for all practical purposes} pragmatic approach).
This loss of the coherence %--usually caused by the interaction with the environment-- 
induces a decreasing of the information on the physical system. For this reason, mixed states represent a crucial tool in quantum decoherence theory \cite{Sch}, \cite{Z}.

As a generalization of the unitary case, it is possible for mixed states also to resort a notion of uncertainty that coherently generalizes the concept considered in the case of pure states. We say that a mixed state $\rho$ represents an uncertain piece of information if its probability value is in the interval $(0,1)$. % (that is, the probability of $\rho$ is neither $0$ nor $1$).
Let us remark that, differently from the unitary case, for an arbitrary density operator $\rho$ the property of being \emph{mixed} is formally equivalent to the fact that 
\begin{equation}\label{eq:mix}
\tr(\rho^2)<1. 
\end{equation}

This corresponds to the non-unitarity of $\rho$: in fact, $\tr(\rho^2)=1$ if and only if $\rho$ is a pure state.
For any density operator $\rho$ on $\otimes^{n}\mathbb C^{d}$, its (normalized) \emph{linear entropy} $SL$, defined as $\frac{d^{n}}{d^{n}-1}(1-\tr(\rho^2))$, provides a measure of its degree of mixedness, or impurity.
%Mixed states arise in situations where the experimenter does not know which particular states are being manipulated.
Clearly, only when $SL(\rho)=0$, $\rho$ represents a maximal piece of information. % that can not be increased by any other possible observation. 
In this case $\rho$ is a {pure state}.

Although uncertainty and mixedness are proper features of a state, they are independent of each other. The following table exemplifies a few simple cases in which they can be told apart.

\begin{table}[htdb]
\caption{}%Degrees of freedom in $\mathfrak D(\tensrn 2)$}
\begin{center}
\begin{tabular}{|c|c|c|}
\hline
 & {\bf Uncertain}& {\bf Not Uncertain}\\\hline
{\bf Mixed}& $\left(
   \begin{matrix}
1-\lambda & 0    \\
0& \lambda
\end{matrix}
\right)$ with $\lambda\in (0,1)$ &
$\left(\begin{matrix}
0 &0    & 0 & 0\\
0&1/2&0&0\\
0&0&0&0\\
0&0&0&1/2
\end{matrix}
\right)$\\\hline
{\bf Not Mixed}&$   \frac12\left(
\begin{matrix}
1 & 1    \\
1&1
\end{matrix}
\right)$ & $\left(
\begin{matrix}
0 & 0    \\
0&1
\end{matrix}
\right)$\\
\hline
\end{tabular}
\end{center}
\label{tab:1}
\end{table}%

\subsection{Degrees of truth}
Quantum computational logics, in the interpretation of Maria Luisa Dalla Chiara, Giampiero Cattaneo and other authors, including the present writers \cite{CDGL}, \cite{CDGL04}, \cite{DGG}, \cite{DGL}, \cite{DGL05}, depart from the usual Birkhoff-von Neumann approach, where meanings of sentences are projection operators, or, equivalently, closed subspaces of a Hilbert space. \\
In this other approach, what really matters are quantum information units: qubits, quregisters, and, more generally, density operators on a given Hilbert space $\otimes^n\mathbb C^2$ \cite{GLSP}. Fuzzy-like structures have appeared in this context which have been extensively studied \cite{DGFL,DGLS,GFLP}. However, as noted in \cite{FLSG}, this fuzzy behavior is mainly due to probabilistic features concerning uncertainty aspects of a state. Namely, the probabilities of a state to be detected either in the ``false" state $\proj02n$, or in the ``true" state $\proj12n$. In fact, the backdrop is an Hilbert space which is a tensor power of the  space $\mbb C^{2}$, where only two possible ``truth-values'' are available: $\ket0$ and $\ket1$. Within this approach, no other truth values are allowed. A state may be uncertain, of course; but it would be uncertain with respect to the true and to the false projector.\\
\noindent Under several aspects, this standpoint may be considered unduly restrictive. For, as mentioned in \cite{BL,MS}, it is absolutely conceivable to encounter cases in which a physical system may collapse into several states. A definite case in point are qutrits \cite{GSSS,T}, where a state $\ket\psi$ may have probabilities $a_{0}, a_{\frac12},a_1$ to be detected in the basis states $\ket0$,  $\ket{\frac{1}{2}}$, $\ket1$, respectively.
Consider, for instance, the states:
\begin{eqnarray*}
\ket\psi=&\frac1{\sqrt2}(\ket0+\ket1) \\
\ket{\psi'}=&\ket{\frac12}
\end{eqnarray*}
in $\mbb C^2$ and $\mbb C^3$, respectively.
%\commento{Fuzzyness come caratteristica dello spazio, dipendente esclusivamente dalla base}
 On the one hand, $\ket \psi$ in the first equation above is an uncertain and pure state.  However, only truth and false probabilities can be associated to $\ket\psi$, since $\ket\psi$ is a state in $\mathbb C^{2}$.
\nel On the other hand, the state $\ket{\psi'}$ is neither uncertain, nor mixed: \emph{it is the state $\ket{\frac12}$}. This other value should not be considered as a superposition of basis elements, because it is on its own another basis element. Indeed, for any $n$, in the tensor power $\tensr n 2$, the only possible truth values are classical; in $\mbb C^{3}$ a new truth-value appears: $\ket{\frac12}$.  \nel The Hilbert space $\mbb C^{4}$ provides a neat example. The spaces $\tensr 2 2$ and $\mathbb C^4$ are clearly isomorphic. Actually they are the same mathematical object. However, from the perspective of quantum computational logics they provide distinct logical semantics.\nel For instance, the vector 
\[
\ket\psi=\ket {01} =\begin{pmatrix}0 \\
                       1 \\
                       0\\
                       0\\

                    \end{pmatrix}
\]
 is a true register in $\otimes^2\mathbb
C^2$, because the last component of the tensor product is $\ket 1$. 
%\[
%\prob1{2}{2}\ket\psi\bra \psi=\tr(\proj122\ket\psi\bra \psi)=\tr\left(\begin{pmatrix}0 &0&0&0\\
%                       0 &1&0&0\\
%                       0 &0&0&0\\
%                       0 &0&0&1\\
%\end{pmatrix}
%\begin{pmatrix}0 &0&0&0\\
%                       0 &1&0&0\\
%                       0 &0&0&0\\
%                       0 &0&0&0\\
%\end{pmatrix}
%\begin{pmatrix}0 &0&0&0\\
%                       0 &1&0&0\\
%                       0 &0&0&0\\
%                       0 &0&0&1\\
%\end{pmatrix}\right)=1
%.
%\]

On the other hand, this can be easily seen to be not the case in $\mbb C^{4}$, since 
\[
\proj14{}=\ket1\bra1=\begin{pmatrix}0 &0&0&0\\
                       0 &0&0&0\\
                       0 &0&0&0\\
                       0 &0&0&1\\
\end{pmatrix}
\]
where $\ket 1\in\mathbb C^4.$

This fact is a consequence of the assumptions we made for the notions of true and false state.
\nel According with Definition \ref{tv}, it is evident that the outcome of a measurement in $\tensr22$ is, in terms of truth and false probabilities, necessarily two-valued, since it is committed to a classical backdrop.\nel This won't be longer the case if a different notion of truth comes into play. For, in $\mbb C^4$, up to the choice of the basis, there are four possible available values: $\ket{0},\ket{\frac13},\ket{\frac23},\ket{1}$.

According with this idea, if we agree on the computational basis, the state 
\[
\ket\psi=\begin{pmatrix}0 \\
                       1 \\
                       0\\
                       0\\
                    \end{pmatrix}
\]
would correspond to the state $\ket{\frac13}$.

This brief observation is expedient to emphasize that the choice of the quantum information units determines the context in which quantum computational logics operate. If we start with a quantum information unit in $\mbb C^2$, then every possible quantum computational logic would rely on a classical two-valued setting. On the other hand, if our choice is $\mbb C^n$, $n>2$, new truth-values come into play.
%
%\commento{Toglierei.}\st{the operation of \emph{increasing the dimension of the space} by using the tensor product or by increasing the dimension of the basis, has a very different logical meaning even though it could bear the same mathematical representation.  Also from a physical point of view, while the tensor product formally represents the interaction among systems, the dimension of the Hilbert space represents the number of possible outcomes of a measurement on a given observable.
%}

From now on, let us focus on the Hilbert space $\mbb C^3$.
{\tiny\begin{table}[htdp]
%\caption{default}
\begin{center}
 \begin{equation*}
  \xymatrix{
\ket{\psi_1}\ar@{-}[rrrr]&&&\ar@{..>}_{\Delta\varepsilon_2}[ddddd]&\varepsilon_0+\Delta\varepsilon_2\\\\\\
{\ket{\psi_{\frac{1}{2}}}} \ar@{-}[rrrr]&\ar@{..>}_{\Delta\varepsilon_1}[dd]&&&\varepsilon_0+\Delta\varepsilon_1\\\\
\ket{\psi_{0}}\ar@{-}[rrrr]&&&&\varepsilon_0
%  &\mbox{\bf S.I. and Subtractive} \ar@{-->}[dl] \ar@{-->}[dr]&\\
%  \mbox{\bf Simple}\ar@{-->}[dd] &&SP_{u}(\mathcal K_{+})\ar@{-->}[dddd]\\\\
%  \mbox{\bf S.I.}\ar@{-->}[dd]&&\\\\
%  \sigma\mbox{\bf -simple}\ar@{<-->}[rr]&&\mbox{\bf Quasi-discriminator}
  }
 \end{equation*}
\end{center}
\vspace{1cm}
\caption{}
\label{tab2}
\end{table}%
}%there is a probability (say, $|a_1|^2$) there is a probability (say, $|a_{\frac{1}{2}}|^2)$ and finally there is a probability (say, $|a_0|^2)$
A concrete physical system that should be necessarily represented by a state in $\mbb C^3$ is depicted in Table \ref{tab2}.
Consider a $3$-levels energy system in an excited state $\ket{\psi_1}$, whose energy is $\varepsilon_0+\Delta\varepsilon_2$. Three events -- with respective probabilities $|a_1|^2$, $|a_2|^2$ and $|a_3|^2$ (related to the respective gap of energy) -- are possible:

\begin{enumerate}
 \item the system remains in the same state $\ket{\psi_1}$;
 \item the system decays in the state $\ket{\psi_{\frac{1}{2}}}$, whose respective energy is $\varepsilon_0+\Delta\varepsilon_1$;
 \item the system collapses to the ground state $\ket{\psi_0}$, whose respective energy is $\varepsilon_0$.
\end{enumerate}

The $3$-levels energy system described above can be formally expressed as a qutrit -- a unit vector in $\mathbb C^3$:
$$\ket{\psi}=a_0\ket{\psi_0}+a_{\frac{1}{2}}\ket{\psi_{\frac{1}{2}}}+a_1\ket{\psi_1}.$$
A natural example comes from the $3$-levels laser in quantum optics \cite{B}, that is a special case of the \emph{population inversion phenomenon} that occurs when a system of atoms exists in a non-equilibrium state such that more atoms are in an excited state than in the ground energy level. 
%The concept of population inversion is a fundamental notion in the workings of standard laser: simply speaking, it is based on the construction of a non-equilibrium physical system where a group of atoms is excited in \red{such a way},\commento{1) In which way?\\
%2) Tutta la frase, da ``simply speaking'' sino al punto, in inglese non ha senso. Per favore, ti chiederei di riformulare.} for a short time interval, all the three (or more) different energy levels are populated by some atoms of the group. 
In this case, the evolution of the state should consider different possible decays from a state to another (different possible energy transactions). The probability of each decay is related with the respective gap of energy. During the dacay process, a laser is emitted and its wavelength depends on the corresponding gap of energy.  

As we will see in the next section, the observations above induce effective consequences in the extension of the definitions of some standard unary quantum logical gates on the Hilbert space $\mathbb C^3$.

\section{Extending the quantum gates}\label{sec:4}
In this section we discuss extensions of several well known quantum gates to the case of qutrits. These constructions exploit the fact that in $\mbb C^3$ -- as well as in $\tensr n 3$-- the new truth value widens the usual behavior of  gates in  $\mbb C^2$ -- as well as in $\tensr n 2$-- along distinct degrees of freedom. In fact, we will see that single gates in $\mbb C^2$ may admit several extensions in the case of qutrits.

\subsection{The negation}
\noindent{\bf Qubit case:}
For any $n\geq 1$, the \emph{negation} on $%
\tensr{n}{2}$ is the unitary operator ${Not}^{%
{(2^n)}}$ such that, for every element $\left\vert
x_{1},...,x_{n}\right\rangle $ of the computational basis $\mathcal{B}^{(2^n)}$%
,%
\begin{equation*}
{Not}^{(2^n)}(\left\vert x_{1},...,x_{n}\right\rangle
)=\left\vert x_{1},...,x_{n-1}\right\rangle \otimes \left\vert
1-x_{n}\right\rangle \text{.}
\end{equation*}

We have that:%
$$
{Not}^{(2^n)}=
\begin{cases}
\sigma _{x} &\text{if } n=1; \\
I^{(n-1)}\otimes \sigma _{x}, &\text{otherwise,}
\end{cases}
$$
where $\sigma_x:=\begin{pmatrix}
  0& 1 \\
  1& 0 \\
\end{pmatrix}$
is the ``first'' Pauli matrix.

{\bf Qutrit case:} Given the usual basis $\mc B^{(3)}=\{\ket 0, \ket{ \frac{1}{2}}, \ket 1\}$ of $ \mathbb C^3$, it is possible to define a negation $Not^{(3)}_{\ket{\frac12}}$ as expected by 
\[
Not^{(3)}_{\ket{\frac12}}\ket{x}=\ket{1-x},
\]
where $x\in\{0,\frac{1}{2},1\}$. We use the subscript $\ket{\frac12}$ to emphasize the fact that $\ket{\frac12}$ is a fixpoint of  $Not^{(3)}_{\ket{\frac12}}$, i.e. $Not^{(3)}_{\ket{\frac12}}\ket{\frac12}=\ket{\frac12}$.\nel
We can easily obtain the matrix form $Not^{(3)}{{_{\ket{\frac{1}{2}}\color{black}}}}=\begin{pmatrix}0 & 0 & 1 \\
0 & 1   &0\\
1  & 0 & 0 \end{pmatrix}$ such that:
\[
Not^{(3)}_{\ket{\frac{1}{2}}}\color{black}\begin{pmatrix}a\\
b\\
c
\end{pmatrix}=\begin{pmatrix}
c\\
b\\
a   
\end{pmatrix}.
\]
This idea can be easily generalized to the other basis states as follows:
\[
Not^{(3)}{{_{\ket0\color{black}}}}=\begin{pmatrix}
1 & 0 & 0 \\
0 & 0   &1\\
0  & 1 & 0
\end{pmatrix}
\text{ and }
Not^{(3)}{{_{\ket1\color{black}}}}=\begin{pmatrix}
0 & 1 & 0 \\
1 & 0   &0\\
0  & 0 & 1
\end{pmatrix}.
\]
Let us remark that, for any $i\in\{0,\frac12,1\}$, $Not^{(3)}_{\ket i}\cdot Not^{(3)}_{\ket i}=I^{(3)}$.

\begin{remark}\label{rem:c2x2-c4}
Note that, given the computational basis $\mc B^{(4)}=\{\ket 0, \ket{ \frac{1}{3}}, \ket{ \frac{2}{3}},\ket 1\}$, and $\mc B^{(2^{2})}=\{\ket{00},\ket{01},\ket{10},\ket{11}\}$ the $Not$-like gates they induce are essentially different. Namely,
\[
Not^{(4)}_{\ket{\frac12}}=
\begin{pmatrix}
0 & 0 & 0 & 1 \\
0 & 0   &1   & 0\\
0  & 1 & 0 & 0 \\
1  & 0 & 0 & 0
\end{pmatrix}
\]

but

\[
Not^{(2^2)}=I^{(2)}\otimes Not^{(2)}
\begin{pmatrix}
0 & 1 & 0 & 0 \\
1 & 0   &0   & 0\\
0  & 0 & 0 & 1 \\
0  & 0 & 1 & 0
\end{pmatrix}.
\]
\end{remark}

\subsection{The Hadamard gate}
{\bf Qubit case:} For any $n\geq 1$
, the \emph{Hadamard gate} on $\tensr{n}{2}$ is the linear operator $H^{(2^n)}$ such that for every element
$\left\vert x_{1},...,x_{n}\right\rangle $ of the computational basis $
\mathcal{B}^{\left( 2^n\right) }$:
\begin{equation*}
H^{(2^n)}(\left\vert x_{1},...,x_{n}\right\rangle
)=
\left\vert x_{1},...,x_{n-1}\right\rangle \otimes \frac{1}{\sqrt{2}}\left(
(-1)^{x_{n}}\left\vert x_{n}\right\rangle +\left\vert 1-x_{n}\right\rangle
\right) \text{.}
\end{equation*}%
We have that
%------------
\begin{equation*}
H^{(2^n)}=
\begin{cases}
H &\text{ if } n=1; \\
I^{n-1}\otimes H, &\text{otherwise,}%
\end{cases}
\end{equation*}
where $H$ is the {\it Hadamard} matrix:
\begin{equation*}
H=\frac{1}{\sqrt{2}}%
\begin{pmatrix}
1 & 1 \\
1 & -1%
\end{pmatrix}.%
\end{equation*}%
The basic property of $H^{(2^n)}$ is that, for any $\left\vert \psi \right\rangle \in \tensr {n}{2}$:
\[
 H^{(2^n)}\left(H^{(2^n)}(\left\vert \psi \right\rangle) \right) =\left\vert \psi \right\rangle.
\]

{\bf Qutrit case:} In \cite{ACB}, the following extension of the Hadamard gate for qutrits is considered

\[
H^{(3)}=\frac{1}{\sqrt3}
\begin{pmatrix}
1 & 1 & 1 \\
1 & \frac{1}{6}(-1+i\sqrt 3)   &-\frac{1}{6}(1+i\sqrt 3)\\
1  & -\frac{1}{6}(1+i\sqrt 3) & \frac{1}{6}(-1+i\sqrt 3)
\end{pmatrix}
\]

as a tool in the framework of distillation protocols for fault tolerant quantum computation- precisely Magic State Distillation.

We state without proof the main properties of $H^{(3)}$:

\begin{lemma}\label{lem:propH3}
\noindent\begin{enumerate}
\item {For any $\ket{\psi}\in\mc B^{(3)}$}, $H^{(3)}\ket\psi=a\ket 0 +b \ket{\frac{1}{2}} + c \ket 1$ s.t $|a|^2=|b|^2=|c|^2=\frac{1}{3}$;
\item $H^{(3)}$ is a genuinely quantum gate;
\item $H^{(3)}\cdot H^{(3)}=Not^{(3)}_{\ket 0} \neq I$;
\item for any density operator
\footnote{The extensive definition of a density operator $\rho$ on $\mathbb C^3$ is provided in the next Section. When a qutrit quantum gate $A$ is applied to a density operator $\rho$ on $\mathbb C^3$, the evolution of $\rho$ is given by: $A\rho A^{\dagger}$. Since no danger of confusion will be impending, for the sake of notational simplicity, from now on we write $A(\rho)$.} 
$\rho$ on $\mbb C^{3}$, $SL(\rho)\neq SL(H^{(3)}(\rho))$.
\end{enumerate} 
\end{lemma}

As shown in the previous lemma, $H^{(3)}$ preserves several desirable features of $H$: it is a genuinely quantum gate that transforms {states in the logical base} $\mc B^{(3)}$ into superposition states with uniformly distributed probabilities.

Along different lines, other possible extensions of the Hadamard gate are the \emph{square root of the identity gates} \cite{GLSP}:
\begin{align*}
\sqrt I_{\ket 0}^{(3)}=1\oplus H_{(\mathbb C^2)}=\frac{\sqrt 2}{2}\begin{pmatrix}
\sqrt 2 & 0 & 0 \\
0 & 1   &1\\
0  & 1 & -1
\end{pmatrix};
& \sqrt I_{\ket {\frac{1}{2}}}^{(3)}=\frac{\sqrt 2}{2}\begin{pmatrix}
1 & 0 & 1 \\
0 & \sqrt 2   &0\\
1  & 0 & -1 
\end{pmatrix};
\\
\sqrt I_{\ket 1}^{(3)}= H_{(\mathbb C^2)}\oplus 1=\frac{\sqrt 2}{2}\begin{pmatrix}
1 & 1 & 0 \\
1 & -1   &0\\
0  & 0 & \sqrt 2 
\end{pmatrix}
&
\end{align*} where $\oplus$ indicates the matrix direct sum.

Some properties of the gates above follow:
\begin{lemma}\label{lem:prp sqrt I} For any $i\in\{0,\frac{1}{2},1\}$:
\noindent\begin{enumerate}
\item $\sqrt I_{\ket i}^{(3)}$ is a \emph{genuinely} quantum qutrit-gate;
\item $\sqrt I_{\ket i}^{(3)}\cdot\sqrt I^{(3)}_{\ket i}=I$;
\item for any density operator $\rho$ on $\mathbb C^3$, $SL(\rho)=SL(\sqrt I^{(3)}_{\ket i}(\rho))$, as in the qubit case;
\item 
\begin{align*}
\sqrt I_{\ket 0}^{(3)}\begin{pmatrix}  
a \\
b\\
c 
\end{pmatrix}=\begin{pmatrix}
a \\
\frac{1}{\sqrt 2}(b+c)\\
\frac{1}{\sqrt 2}(b-c) \end{pmatrix};\; &
\sqrt I_{\ket {\frac{1}{2}}}^{(3)}\begin{pmatrix}
a\\
b\\
c
\end{pmatrix}=\begin{pmatrix}  
\frac{1}{\sqrt 2}(a+c) \\
b\\
\frac{1}{\sqrt 2}(a-c) 
\end{pmatrix};\;
&
 \sqrt I_{\ket 1}^{(3)}\begin{pmatrix}
a\\
b\\
c
\end{pmatrix}=\begin{pmatrix}
\frac{1}{\sqrt 2}(a+b) \\
\frac{1}{\sqrt 2}(a-b)\\
c
\end{pmatrix}.
\end{align*}
\end{enumerate}
\end{lemma}
 Let us remark that, in the Hilbert space $\mathbb C^3$ the Hadamard gate $H^{(3)}$ does not behave as a square root of the identity. Instead, $\sqrt I^{(3)}_{\ket i}$ is a square root of identity for any $i\in\{0,\frac{1}{2},1\}.$
\subsection{The Square-Root of the Negation}
{\bf Qubit case:} For any $n\geq 1$, the square
root of the negation \cite{CDGP04} on $\tensr n 2$ is the
unitary operator $\sqrt{{Not}}^{{(2^n)}}$ such that, for every
element $\left\vert x_{1},...,x_{n}\right\rangle $ of the computational
basis $\mathcal{B}^{(2^n)}$,%
\begin{equation*}
\sqrt{{Not}}^{{(2^n)}}(\left\vert x_{1},...,x_{n}\right\rangle
)=\left\vert x_{1},...,x_{n-1}\right\rangle \otimes \frac{1}{2}\left(
(1+i)\left\vert x_{n}\right\rangle +(1-i)\left\vert 1-x_{n}\right\rangle
\right) \text{.}
\end{equation*}%
The basic property of $\sqrt{{Not}}^{{(2^n)}}$ is the following:
for any $\left\vert \psi \right\rangle \in\tensr n 2$,%
\begin{equation*}
\sqrt{{Not}}^{{(2^n)}}\left(\sqrt{{Not}}^{{(2^n)}}\left( \left\vert \psi \right\rangle \right) \right) ={Not}^{{
(2^n)}}\left( \left\vert \psi \right\rangle \right) \text{.}
\end{equation*}%
From a logical point of view, therefore, the square root of the negation can be
regarded as a kind of ``tentative partial negation" that transforms precise
pieces of information into maximally uncertain ones. For, we have
\begin{equation*}
\prob1{2}{}(\sqrt{{Not}}^{{(2)}}(\left\vert 0\right\rangle ))=%
\frac{1}{2}=\prob1{2}{}(\sqrt{\mathtt{Not}}^{\mathtt{(1)}}(\left\vert
1\right\rangle ))\text{.}
\end{equation*}%
As noticed in \cite[Lemma 17.1.11] {DGG}, this gate possesses no Boolean counterpart.
\nel {\bf Qutrit case:} As in the case of the gate $Not^{(3)}_{\ket{i}}$, $i\in\{0,\frac{1}2, 1\}$, once we enter in the qutrit world, several possible widening of $\sqrt{{Not}}^{(2)}$ are available. Namely, given the usual basis $\mc B^{(3)}$ of $ \mathbb C^3$, it is possible to define the gates $\sqrt {Not}_{\ket 0}^{(3)}$, $\sqrt {Not}_{\ket {\frac12}}^{(3)}$, and $\sqrt {Not}_{\ket 1}^{(3)}$ as follows:
\begin{eqnarray*}
\sqrt {Not}_{\ket 0}^{(3)}&=&1\oplus \sqrt {Not}^{(2)}=\frac{1}{2}\begin{pmatrix} 2 & 0 & 0 \\
                      0 & 1+i   &1-i\\
      
                        0  & 1-i &1+i 
                     \end{pmatrix};\\ 
\sqrt {Not}_{\ket {\frac{1}{2}}}^{(3)}&=&\frac{1}{2}\begin{pmatrix}1+i & 0 & 1-i \\
                      0 &  2   &0\\
      
                        1-i & 0 & 1+i 
                     \end{pmatrix};\\
\sqrt {Not}_{\ket 1}^{(3)}&=& \sqrt{Not}^{(3)}\oplus 1=\frac{1}{2}\begin{pmatrix}1+i & 1-i & 0 \\
                      1-i & 1+i   &0\\
      
                        0  & 0 & 2 
                     \end{pmatrix}.
\end{eqnarray*}

Similarly to $\sqrt I^{(3)}_{\ket i}$, it can be seen that the gates $\sqrt {Not}_{\ket i}^{(3)}$ also, act ``locally'' as genuinely quantum gates on two of the base vectors, leaving unchanged $\ket0, \ket{\frac12}, \ket1$, respectively:

\begin{eqnarray*}
\sqrt {Not}_{\ket 0}^{(3)}\begin{pmatrix}  a \\
                      b\\
      
                        c
                     \end{pmatrix}&=&\begin{pmatrix}  a \\
                      \frac{1}{2}[(1+i)b+(1-i)c]\\
      
                        \frac{1}{2}[(1-i)b+(1+i)c] 
                     \end{pmatrix} ;\\
\sqrt {Not}_{\ket {\frac{1}{2}}}^{(3)}\begin{pmatrix}  a \\
                      b\\
      
                        c 
                     \end{pmatrix}&=&\begin{pmatrix}  \frac{1}{2}[(1+i)a+(1-i)c] \\
                     b\\
      
                        \frac{1}{2}[(1-i)a+(1+i)c]
                     \end{pmatrix};\\
\sqrt {Not}_{\ket 1}^{(3)}\begin{pmatrix}  a \\
                      b\\
      
                        c 
                     \end{pmatrix}&=&\begin{pmatrix}  \frac{1}{2}[(1+i)a+(1-i)b] \\
                      \frac{1}{2}[(1-i)a+(1+i)b]\\
      
                        c 
                     \end{pmatrix}.
\end{eqnarray*}

Some interesting properties of the gates above are summarized in the following: 

\begin{lemma}\label{lem:sqrtnot prop}
For any $i\in\{0,\frac{1}{2},1\}$, 
\begin{enumerate}
\item $\sqrt {Not}_{\ket i}^{(3)}$ is a \emph{genuinely} quantum qutrit-gate;
\item $\sqrt {Not}_{\ket i}^{(3)}\cdot\sqrt {Not}_{\ket i}^{(3)}=Not^{(3)}_{\ket i}$;
\item for any density operator $\rho$ on $\mathbb C^3$, $ SL(\rho)=SL(\sqrt {Not}^{(3)}_{\ket i}(\rho))$ 
(as in the qubit case).
\end{enumerate}
 \end{lemma}

\section{Effect caracterization of density operators on $\mathbb C^3$}\label{sec:5}
As mentioned in \ref{M}, when an interaction between a system and the environment comes into play, the state of the system is represented by a qumix. 
It is well known that Pauli matrices $\sigma_1,\sigma_2,\sigma_3$
% \begin{equation*}
%  \sigma_1 =
%  \left[\begin{array}{cc}
%    0 & 1 \\
%    1 & 0
%  \end{array}\right],
%  \quad
%  \sigma_2 =
%  \left[\begin{array}{cc}
%    0 & -i \\
%    i & 0
%  \end{array}\right],
%  \quad
%  \sigma_3 =
%  \left[\begin{array}{cc}
%    1 & 0 \\
%    0 & -1
%  \end{array}\right]
%\end{equation*}
and $I$ form a basis for the set of density operators on ${\mathbb{C}}^2$, so that an arbitrary density operator $\rho$ in ${\mathbb{C}}^2$ may be represented
as $$\rho=\frac{1}{2}(I+r_1\sigma_1+r_2\sigma_2+r_3\sigma_3)$$ where $r_1,r_2$ and $r_3$ are real numbers such that $r_1^2+r_2^2+r_3^2\le 1$. The vector $(r_1,r_2,r_3)$ represents the uniquely determined point in the Bloch sphere  associated to $\rho$: the \emph{Bloch-vector} of $\rho$. There is a one-to-one correspondence between the space of the lenght-$1$ vectors in $\mathbb R^3$ and the space of the density operators in $\mathbb C^2$.

A representation of this sort can be obtained for any $n$-dimensional Hilbert space through generalized Pauli-matrices. In particular, in $\mathbb C^3$, Pauli matrices are generalized by Gell-Mann matrices \cite{SM}.

%
% \begin{equation*}
%  \lambda_1 =
%  \left[\begin{array}{ccc}
%    0 & 1 & 0\\
%    1 & 0 & 0\\
%    0 & 0 & 0
%  \end{array}\right],
%  \quad
%  \lambda_2 =
%  \left[\begin{array}{ccc}
%    0 & -i & 0\\
%    i & 0 & 0\\
%    0 & 0 & 0
%  \end{array}\right],
%  \quad
%  \lambda_3 =
%  \left[\begin{array}{ccc}
%    1 & 0   & 0\\
%    0 & -1 & 0\\
%    0 & 0   & 0
%  \end{array}\right],
%\quad
%  \lambda_4 =
%  \left[\begin{array}{ccc}
%    0 & 0 & 1\\
%    0 & 0 & 0\\
%    1 & 0 & 0
%  \end{array}\right],
%\end{equation*}
%
% \begin{equation*}
%  \lambda_5 =
%  \left[\begin{array}{ccc}
%    0 & 0 & -i\\
%    0 & 0 & 0\\
%    i & 0 & 0
%  \end{array}\right],
%  \quad
%  \lambda_6 =
%  \left[\begin{array}{ccc}
%    0 & 0 & 0\\
%    0 & 0 & 1\\
%    0 & 1 & 0
%  \end{array}\right],
%  \quad
%  \lambda_7 =
%  \left[\begin{array}{ccc}
%    0 & 0   & 0\\
%    0 & 0 & -i\\
%    0 & i   & 0
%  \end{array}\right],
%\quad
%  \lambda_8 =\frac{1}{\sqrt3}
%  \left[\begin{array}{ccc}
%    1 & 0 & 0\\
%    0 & 1 & 0\\
%    0 & 0 & -2
%  \end{array}\right].
%\end{equation*}

It can be seen that any density operator $\rho$ in $\mathbb C^3$ \cite{GSSS} can be written as
\begin{equation}\label{eq:rho}
\rho=\frac{1}{3}(I+\sqrt3\sum_{i=1}^8 r_i\lambda_i),
%=\frac{1}{3}\left[\begin{array}{ccc}
%    \sqrt3r_3+r_8+1 & \sqrt3(r_1-ir_2)   & \sqrt3(r_4-ir_5)\\
%    \sqrt3(r_1+ir_2) & -\sqrt3r_3+r_8+1 & \sqrt3(r_6-ir_7)\\
%    \sqrt3(r_4+ir_5) & \sqrt3(r_6+ir_7)   & 1-2r_8
%  \end{array}\right] 
\end{equation}

where $r_i$ are real numbers such that $\sum_{i=1}^8|r_i|^2=1.$

Let us observe that lenght-$1$ vectors in $\mathbb R^8$ and the space of the density operators in $\mathbb C^3$ are not in one-to-one correspondence.
Indeed, if $r_i=1$ for $i=8$ and $r_i=0$ otherwise, then the eigenvalues of the corresponding operator $\rho$ in Equation \eqref{eq:rho} would be $\{\frac{2}{3},\frac{2}{3},-\frac{1}{3}\}$, against the positivity requirement.%, and so $\rho$ is not a density operator.

In this section, we propose an alternative characterization of density operators in $\mbb C^{3}$ through the notion of \emph{effect probability}.
\begin{definition}
Let $\mbb H$ be a complex Hilbert space that represents the state space of a quantum system S. The set of \emph{effects} $E(\mbb H)$ for $S$ is the set of operators
\[
\{A: \forall\ket{\psi}\in \mbb H, 0\le\langle{\psi}\ket{A\psi}\le1 \}.
\]
\end{definition}
Following \cite{G}, an effect represents a yes-no measurements (for example a measurement on a Stern-Gerlach apparatus, whose outputs can only be spin up or spin down \cite[I.1.2]{BGL}) that may be \emph{unsharp} \cite{GG}: sharp measurements are mere idealization, impossible in practice, where measurements are always imprecise to some degree. This unsharpness arises from the interaction between the system and the environment. For example, consider a geiger or a photon counter performing a position measurement on a one-particle quantum system. If the system is completely isolated from the environment and the detector is perfectly accurate, then it clicks if and only if the particle is detected within a certain sensitivity domain $B\subset R^3$. However, this situation is a limit condition, not encompassing real situations, where the system interacts with the environment and the measurement device is not perfectly accurate. In this case, an adequate calibration experiment classifies the confidence that the apparatus clicks when the particle is in $B$ as a real value in $(0,1)$. According to \cite{G}, this can be considered an example of a \emph{yes-not measurement} quantum event (the particle is/is not in a certain sensitivity domain $B$) to whose outcomes a probability value can be assigned.

%, to which a probability value some kind of \emph{unsharpness} (or fuzziness) due to the not-perfect accuracy of the measurement device. 

Phenomena of this sort motivated the development of POVM (positive operator valued measure) theory, that generalises standard PVM (positive valued measure) theory \cite[\S 1.3]{BGL}. POVM relies on the fact that the physical reality is described as it emerges when investigated by measuring processes, which are to be considered, themselves, as physical processes. 
Perhaps, the striking difference between PVM and POVM is that, while PVM considers only sharp observables, represented by hermitian operators, and projective measurements, POVM widens this setting to unsharp observables, represented by effects, to which effect-valued measurements, that generalize within this context  projective measurements, are associated \cite[p.6]{BGL}. 
Indeed, to any effect valued measurement a probability measurement (effect probability) is naturally associated by the Born rule \cite[(1.21)]{BGL}:
\begin{equation}\label{eq: eff prob}
\mathtt{Pr}_{(E)}(\rho)=\mathtt{tr}(E\rho), 
\end{equation}

where $E$ and $\rho$ are an effect and a density operator on an Hilbert space $\mbb H$, respectively.

In the case of $\mbb C^{3}$, the notion of effect probability turns out to be particularly expedient for our purposes if $E$ in \eqref{eq: eff prob} is the effect:
\[
E=\left(\begin{array}{ccc}
    0 & 0   & 0\\
    0 & \frac{1}{2} & 0\\
    0 & 0 & 1
\end{array}\right).
\]
In fact, using this particular effect probability we can define a three-valued quantum computational logic on $\mbb C^{3}$ that consistently generalizes two-valued quantum computational logic on $\mbb C^{3}$. Indeed, for any $\ket i$ in the computational basis, $\mathtt{Pr}_{(E)}(\ket i)=i$.
Moreover, $\mathtt{Pr}_{(E)}$ is crucial in proving the following:

%
%\begin{definition}
%For any qutrit-density operator $\rho\in\mathfrak D(\mathbb C^3)$ we can define the \emph{effect-probability} as:
%$$P_E(\rho)=Tr(E\rho),$$
%where $E=\left[\begin{array}{ccc}
%    0 & 0   & 0\\
%    0 & \frac{1}{2} & 0\\
%    0 & 0 & 1
%  \end{array}\right].$
%\end{definition}

%\red{Based on what discussed above, the effect-probability can be interpreted as the probability that the outcome of a yes-not unsharp measurement is ``yes".
%}\commento{\red{Come mai?}}
%\commento{Questo esempio mi sembra una banalit\`a, se non motivato.}
%\blue{As an example, if we consider the probabilistically uniformed distribuited qutrit-density operator $\rho=\ket{\psi}\bra{\psi}$, where $\ket{\psi}=\frac{1}{\sqrt3}(\ket0+\ket{\frac{1}{2}}+\ket1)$, it is straightforward to see that $P_E(\rho)=\frac{1}{2}.$}

%
%\begin{figure}
%\begin{center}
%\includegraphics[scale=0.3]{unsharp.jpg}
%\caption{}
%\end{center}
%\end{figure}
%
%What we introduced about the difference between PVM and POVM can be briefly summarized in Figure 1. 
%Even if the debate on this topic is very  large \cite{G,GG}, actually it is out of the purposes of this article; but the definition of effect-probability is crucial for the following proposition.\commento{Puoi mettere dei riferimenti pi\`u recenti di \cite{G,GG}, a parte il libro di Busch?}

\begin{proposition}\label{prop: char}
Let $\rho$ be a density operator in $\mathbb C^3$ and  consider the following set of gates: $A=\{I, Not^{(3)}_{\ket 0}, \sqrt I^{(3)}_{\ket i}, \sqrt{Not^{(3)}}_{\ket i}\}_{i\in\{0,\frac{1}{2},1\}}$. Then $\rho$ is uniquely characterized by the following equations:

\begin{enumerate}
\item $\mathtt{Pr}_{(E)}(\rho)=\frac{1}{6}(3-\sqrt 3 r_3-3r_8)$
\item $\mathtt{Pr}_{(E)}(Not^{(3)}_{\ket 0}\rho)=\frac{1}{2}-\frac{r_3}{\sqrt 3}$
\item $\mathtt{Pr}_{(E)}(\sqrt I^{(3)}_{\ket 0} \rho)=\frac{1}{12}(6-3\sqrt3r_3-2\sqrt3r_6-3r_8)$
\item $\mathtt{Pr}_{(E)}(\sqrt I^{(3)}_{\ket{\frac{1}{2}}}\rho)=\frac{1}{2}-\frac{r_4}{\sqrt 3}$
\item $\mathtt{Pr}_{(E)}(\sqrt I^{(3)}_{\ket1}\rho)=\frac{1}{6}(3-\sqrt3r_1-3r_8)$
\item $\mathtt{Pr}_{(E)}(\sqrt{Not}^{(3)}_{\ket 0} \rho)=\frac{1}{12}(6-3\sqrt 3r_3-2\sqrt3r_7-3r_8)$
\item$\mathtt{Pr}_{(E)}(\sqrt{Not}^{(3)}_{\ket{\frac{1}{2}}}\rho)=\frac{1}{2}-\frac{r_5}{\sqrt 3}$
\item$\mathtt{Pr}_{(E)}(\sqrt{Not}^{(3)}_{\ket 1}\rho)=\frac{1}{6}(3-\sqrt3r_2-3r_8)$

\end{enumerate}
\begin{proof}
Simply notice that the linearly independent Equations (1)-(8)  uniquely characterize the Bloch vector associated to $\rho$.
\end{proof}
\end{proposition}

\section{Conclusions}\label{sec:6}
The aim of this paper was foundational. In particular, we tried to spell out how the concepts of uncertainty and truth-degree meet new degrees of freedom in the framework of quantum computational logics.
A natural future development of the ideas in this paper would include:
\begin{itemize}
\item a study of binary qutrit-gates;
\item a generalization of Proposition \ref{prop: char} to an arbitrary Hilbert space.
\end{itemize}

%\commento{La bibliografia conteneva/contiene un caterva di errori. Sia stilistici: tutte le entrate devono seguire la stessa forma, anche virgole, maiuscole, punti, spazi, ordini, corsivi etc. sono importanti, secondo me. Sia di riferimento. Per esempio:
%non \`e Zurek W. H.,  ``Decoherence and the transition from quantum to classical—revisited". In: \emph{Quantum Decoherence}, p. 1-31, Birkhäuser Basel, 2007.
%Ma, Zurek W. H.,  ``Decoherence and the transition from quantum to classical -- revisited", \emph{Progress in Mathematical Physics}, 48, pp. 1--31, 2007. Allo stesso modo tante altre.} 


\begin{thebibliography}{99}
\bibitem{ACB} Anwar H., Campbell E.T., Browne D.E., ``Qutrit magic state distillation'', \emph{arXiv:1202.2326v2}, 2012.%, 063006, DOI 10.1088/1367-2630/14/6/063006.

\bibitem{BL} Bertini C., Leporini R., ``Quantum computational finite-valued logics'', \emph{International Journal of Quantum Information}, 5, pp. 641-665, 2007.

\bibitem{B} Broers B., van Linden van den Heuvell H. B., Noordam L. D., ``Efficient population transfer in a $3$-level ladder system by frequency-swept ultrashort laser-pulses", \emph{Physical Review Letters}, 69, 14, pp. 2062--2065, 1992.

\bibitem{BGL} Bush P., Grabowski M., Lahti P.J., {\it Operational Quantum Physics}, Springer-Verlag, Berlin and Heidelberg, 1995.

\bibitem{CDGL} Cattaneo G., Dalla Chiara M.L., Giuntini R., Leporini R., ``An unsharp logic from quantum computation'', \emph{International Journal of Theoretical Physics}, 43, 7-8, pp. 1803-1817, 2004.

\bibitem{CDGL04} Cattaneo G., Dalla Chiara M.L., Giuntini R., Leporini R., ``Quantum computational structures'', \emph{Mathematica Slovaca}, 54, pp. 87-108, 2004.

\bibitem{CDGP04} Cattaneo G., Dalla Chiara M.L., Giuntini R., Paoli F., ``Quantum logic and nonclassical logics'', in Engesser K., Gabbay D. M., Lehmann D. (Eds.), \emph{Handbook of Quantum Logic and Quantum Structures}, Elsevier, Amsterdam, pp. 127-235, 2009.

\bibitem{CM} Chernoff H., Moses L.E., \emph{Elementary Decision Theory}, Wiley, New York, 2012.

\bibitem{DGG} Dalla Chiara M. L., Giuntini R., Greechie R., \emph{Reasoning in Quantum Theory}, Kluwer, Dordrecht, 2004.

\bibitem{DGL} Dalla Chiara M. L.,
Giuntini R., Leporini R., ``Quantum computational logics: A
survey", in Hendricks V. F., Malinowski J. (Eds.), \emph{Trends in
Logic: 50 years of Studia Logica}, Kluwer, Dordrecht, pp.
229-271, 2003.

%\bibitem{Dalla Chiara et al. 2005} Dalla Chiara M. L.,
%Giuntini R., Leporini R., ``Logics from quantum computation",
%\emph{International Journal of Quantum Information}, 3, 2, pp. 293-337, 2005.

\bibitem{DGL05} Dalla Chiara M. L., Giuntini R., Leporini R., ``Logics from quantum computation'', \emph{International Journal of Quantum Information}, 3, 2, pp. 293-337, 2005. 

\bibitem{DGFL} Dalla Chiara M. L., Giuntini R., Freytes H., Ledda A., Sergioli G., \textquotedblleft The algebraic structure of an approximately
universal system of quantum computational gates\textquotedblright , \textit{%
Foundations of Physics}, 39, 6, pp 559-572, 2009.

\bibitem{DGLS}  Dalla Chiara M. L., Giuntini R., Ledda A., Sergioli G., \textquotedblleft The Toffoli-Hadamard gate system: an algebraic approach'', \textit{Journal of Philosophical Logic}, 42, 3, pp. 467-481, 2013.

%\bibitem{DGL} Dalla Chiara M. L., Giuntini R., Leporini R., ``Quantum computational logics: a survey'', in Hendricks V. F., Malinowski J. (Eds.), \emph{Trends in Logic: 50 years of Studia Logica}, Kluwer, Dordrecht, Vol.21, pp. 229-271, 2003.


\bibitem{DGS} Dalla Chiara M. L., Giuntini R., Sergioli G., ``Probability in quantum computation and in quantum computational logics. A survey'', \emph{Mathematical Structures in Computer Science}, 24, 03, pp. 1-14, 2014.

\bibitem{DD} Der Kiureghiana A., Ditlevsen O., ``Aleatory or epistemic? Does it matter?'', \emph{Structural Safety},  31, 2, pp. 105-112, 2009.

\bibitem{FLSG} Freytes H., Ledda A., Sergioli G., Giuntini R., \textquotedblleft Probabilistic logics in quantum computation'', in Andersen et al. (Eds.) {\it New Challenges to the Philosophy of Science}, 2013, %ISBN 978-94-007-5844-5, 
pp. 49-57, 2013.

\bibitem{Fi} Fine K., ``Vagueness, truth and logic",
\emph{Synthese}, 30, 3-4, pp. 265-300, 1975.

\bibitem{GFLP} Giuntini R., Freytes H., Ledda A., Paoli F., 
\textquotedblleft A discriminator variety of G\"odel algebras with operators
arising in quantum computation'', \textit{Fuzzy Sets and Systems}, 160, 8, 
pp. 1082-1098, 2009.

\bibitem{GG} Giuntini R., Greulin H., \textquotedblleft Toward a Formal Language for Unsharp Properties'', \textit{Foundations of Physics}, 19, 7, pp. 931-945, 1989.

\bibitem{GLSP} Giuntini R., Ledda A., Sergioli G., Paoli F.,
\textquotedblleft Some generalisations of fuzzy structures arising in quantum computational logic'', \textit{International Journal of General Systems}, 40, 1, pp. 61-83, 2011.

\bibitem{Goguen 69} Goguen J.A., ``The logic of inexact concepts", \emph{%
Synthese}, 19, 3-4, pp. 325-373, 1969.


\bibitem{GSSS} Goyal S.K., Simon N.B., Singh R., Simon S.,\textquotedblleft Geometry of the generalized Bloch sphere for qutrit'', \textit{arXiv:1111.4427}, 2011.


\bibitem{G} Gudder S., \textquotedblleft Sharp and unsharp quantum effects'', \emph{Advances in Applied Mathematics}, 20, 2, pp. 169-187, 1998.

\bibitem{Ha} H\`ajek P., \emph{Metamathematics of Fuzzy logic}, Kluwer, Dordrecht, 1998.

\bibitem{Keefe2000} Keefe R., Smith P. \emph{Vagueness: A reader}, MA:MIT press, Cambridge, 1996.

\bibitem{Lind} Lindley D.V., \emph{Understanding Uncertainty},  Wiley-Sons, New York, 2006.

\bibitem{Matt} Matthies H. G., ``Quantifying uncertainty: modern computational representation of probability and applications'', \emph{Extreme Man-Made and Natural Hazards in Dynamics of Structures}, NATO Security through Science Series, pp. 105-135, 2007.

\bibitem{MS} Muthukrishnan A., Stroud C.R. Jr., \textquotedblleft Multi-valued logic gates for quantum computation'', \emph{Physical Review A}, 62, 5, pp. 0523091--8, 2000.

\bibitem{PC} Pat\'e-Cornell M. E., ``Uncertainties in risk analysis: six levels of treatment'', \emph{Reliability Engineering and System Safety}, 54, 2-3, pp. 95-111, 1999.

\bibitem{SM} Schlienz J., Mahler G., \textquotedblleft Description of entanglement'', \emph{Physical Review A}, 52, 6, pp.4396-4404, 1995.

\bibitem{Sch} Schlosshauer M., \emph{Decoherence and the Quantum-to-Classical Transition}, 1st ed., Springer, Berlin-Heidelberg, 2007.

\bibitem{T} Tarasov V.E.  \textquotedblleft Quantum computation by quantum operations on mixed states'', \textit{arXiv:quant-ph/0201033}, 2002.

\bibitem{TB} Timpson C. G., Brown H. R., ``Proper and improper separability", \emph{The International Journal of Quantum Information}, 3, pp. 679-690, 2005.

\bibitem{Williamson 94} Williamson T., \emph{Vagueness},
Routledge, London, 2002.

\bibitem{Z} Zurek W. H.,  ``Decoherence and the transition from quantum to classical -- revisited", \emph{Quantum Decoherence. Progress in Mathematical Physics}, 48, pp. 1-31, 2007.


\end{thebibliography}
\end{document}